\documentclass[superscriptaddress,twocolumn,prl,aps,10pt]{revtex4-1}

\usepackage{amsfonts,amssymb}
\usepackage[sumlimits,intlimits]{amsmath}
\usepackage{graphics}
\usepackage{graphicx}
\usepackage{mathrsfs}
\usepackage{textcomp}
\usepackage{verbatim}
\usepackage{hyperref}    % links
\usepackage{bm}          % bold math letters

\newcommand{\be}{\begin{equation}}
\newcommand{\ee}{\end{equation}}
\newcommand{\tp}{t}
\newcommand{\e}{\varepsilon}
\newcommand{\rr}{\vec{r}}
\newcommand{\kk}{\vec{k}}
\newcommand{\qq}{\vec{q}}
\newcommand{\pt}{\widetilde{\psi}}

\begin{document}

\title{Bound state energy of a Coulomb impurity in gapped bilayer graphene: \\``Hydrogen atom with a Mexican hat"}

\author{Brian Skinner}
\affiliation{School of Physics and Astronomy, University of Minnesota, Minneapolis, MN  55455, USA}
\affiliation{Fine Theoretical Physics Institute, University of Minnesota, Minneapolis, MN 55455, USA}
\affiliation{Materials Science Division, Argonne National Laboratory, Argonne, IL 60439, USA}
\author{B. I. Shklovskii}
\affiliation{School of Physics and Astronomy, University of Minnesota, Minneapolis, MN  55455, USA}
\affiliation{Fine Theoretical Physics Institute, University of Minnesota, Minneapolis, MN 55455, USA}
\author{M. B. Voloshin}
\affiliation{School of Physics and Astronomy, University of Minnesota, Minneapolis, MN  55455, USA}
\affiliation{Fine Theoretical Physics Institute, University of Minnesota, Minneapolis, MN 55455, USA}
\affiliation{Institute  of Theoretical and Experimental Physics, Moscow, 117218, Russia}

\date{\today}

\begin{abstract}

Application of a perpendicular electric field induces a band gap in bilayer graphene, and it also creates a ``Mexican hat" structure in the dispersion relation.  This structure has unusual implications for the hydrogen-like bound state of an electron to a Coulomb impurity.  We calculate the ground state energy of this hydrogen-like state as a function of the applied interlayer voltage and the effective fine structure constant.  Unlike in the conventional hydrogen atom, the resulting wavefunction has many nodes even in the ground state.  Further, the electron state undergoes ``atomic collapse" into the Dirac continuum both at small and large voltage.

\end{abstract}

\maketitle

One of the primary allures of bilayer graphene (BLG) is its promise of a tunable band gap.  In BLG, the energy gap between the conduction and valence bands can be continuously tuned from less than a few meV to nearly $300$\,meV by the application of a perpendicular electric field $\mathcal{E}$ \cite{McCann2006lld, McCann2006agi, Castro2007bbg, Ohta2006ces, Zhang2009dow}.  The corresponding applied potential difference $V  = \mathcal{E} c_0$ between the two layers, where $c_0 \approx 3.4$\,\AA\ is the interlayer spacing, determines the band gap $\Delta$ according to the relation \cite{McCann2013epb}
\be 
\Delta = \frac{V \tp}{\sqrt{\tp^2 + V^2}},
\label{eq:Delta}
\ee 
where $\tp \approx 300$\,meV is the interlayer coupling energy.  Such a tunable band gap opens up possibilities for new transistor devices.

Unfortunately, achieving a true insulating state in BLG is difficult experimentally owing to the presence of disorder, which creates electron states in the gap \cite{Miyazaki2010idc, Zou2010tig, Taychatanapat2010eti, Oostinga2008gii, Jing2010qta, Yan2010cti, Young2012ecl}.  Such mid-gap states shunt the conductivity at finite temperature, providing pathways for hopping conduction \cite{Zou2010tig, Taychatanapat2010eti, Oostinga2008gii, Jing2010qta, Yan2010cti, Rossi2011ies} and producing a significant density of states that is visible in the electronic compressibility \cite{Young2012ecl}.  Future development of BLG-based devices thus requires an understanding of disorder-induced mid-gap energy states.  

So far theoretical studies of these disorder states have focused primarily on models of uncorrelated, short-range disorder \cite{Young2011cgb, Abergel2012cmm, Mkhitaryan2008dit}.  In this paper we focus on a more basic problem: the energy level produced by a single, isolated Coulomb impurity.  We center our discussion primarily around the case of a co-planar impurity, which for definiteness we take to be positive.  In this sense the problem we consider is similar to the problem of finding the ground state of the two-dimensional (2D) hydrogen atom \cite{Stern1967pss, Chaplik1972cii, Ho1998dei}.

Unlike for the hydrogen atom, however, our problem cannot be solved by direct application of the conventional massive Dirac equation (as in gapped monolayer graphene \cite{Pereira2008sci}), because the electron kinetic energy in gapped BLG has a fundamentally different dependence on momentum.  In particular, at finite $V$ the dispersion relation $\e(\vec{k})$ has a ``Mexican hat" shape \cite{McCann2006lld}, as illustrated in Figs.\ 1a--c, and the band edge is located at a ring in $k$-space defined by $k = |\vec{k}| = k_0$ with \cite{McCann2013epb}
\be 
k_0 = \frac{V}{2 \hbar v} \sqrt{ \frac{2 \tp^2 + V^2}{\tp^2 + V^2} }.
\label{eq:k0}
\ee 
(Here, $\hbar$ is the reduced Planck constant and $v \approx 1.0 \times 10^6$\,m/s is the Dirac velocity in graphene.)  The primary goal of this paper is to determine the binding energy of an electron with such a dispersion relation to the Coulomb center, or in other words to calculate the ground state of the ``hydrogen atom with a Mexican hat."

\begin{figure}[htb]
\centering
\includegraphics[width=0.48 \textwidth]{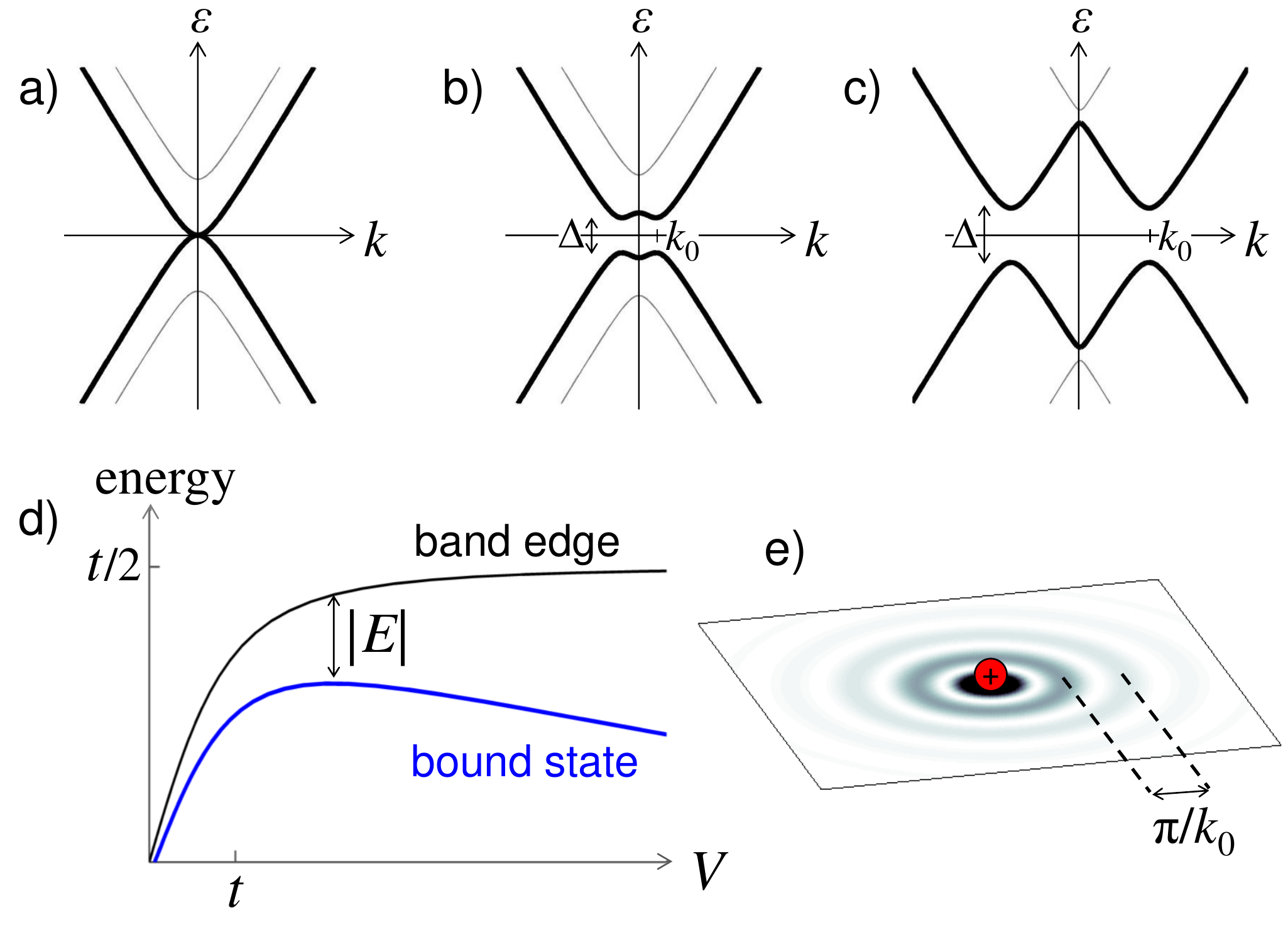}
\caption{(Color online) a)--c) The dispersion relation $\e(k)$ of BLG, shown at a) $V = 0$, b) $V < \tp$, and c) $V > \tp$.  Thick black lines show the valence and conduction bands, while thin gray lines show the outer bands, which are not relevant for this work.  d) The band edge $\Delta/2$ and the impurity state energy $\Delta/2 + E$, plotted as a function of voltage for $Z \alpha = 0.1$.  e) A schematic picture of the corresponding electron density $|\psi(r)|^2$, which has a strongly oscillating component with wave vector $2k_0$ and an exponentially decaying envelope. }
\label{fig:results}
\end{figure}

In general, the ground state energy $E < 0$ of the electron, measured relative to the conduction band edge, depends both on the voltage $V$ between the layers and on the strength of the Coulomb interaction.  The latter can be characterized by the dimensionless parameter $Z \alpha$, where $Ze$ is the charge of the Coulomb impurity and $\alpha = e^2/\kappa \hbar v \approx 2.2/\kappa$ (in Gaussian units) is the effective fine structure constant, with $\kappa$ being the dielectric constant of the media surrounding the BLG.  An exact solution for $E$ is difficult, since it requires one to solve an eigenvalue problem with four coupled linear differential equations (associated with BLG's four atoms per unit cell) \cite{McCann2013epb}.  In this paper, we therefore focus our attention on the simple case where the electron bound state is sufficiently shallow that one can use a single-band approximation, $|E| \ll \Delta$.  As we show below, this corresponds to $Z \alpha \ll 1$ and $(Z \alpha)^2 \ll V/\tp \ll Z \alpha \exp[1/Z \alpha]$.

Our main result for this problem is illustrated in Figs.\ 1d--e.  The energy of the electron bound to the Coulomb impurity is given by
\be 
E \simeq -\frac{2 Z e^2}{\kappa a} \ln^2 (k_0 a),
\label{eq:E}
\ee 
where $a$ is an effective ``Bohr radius" $a = \kappa \hbar^2/m^* Z e^2$ and $m^*$ is an effective mass determined by the curvature of the bottom of the band (defined below).  
The corresponding ground state wavefunction is described by
\be 
\psi(r) \simeq A J_0(k_0r) \exp \left[ - \frac{2 r}{a} \ln (k_0 a) \right],
\label{eq:psi}
\ee 
where $A$ is a normalization constant.  The electron density $|\psi(r)|^2$ is shown schematically in Fig.\ \ref{fig:results}e.  Notably, unlike in conventional quantum mechanics problems, the wavefunction in this case has many nodes of density even in the ground state as a result of the degeneracy of the conduction band minimum.

Within the single band approximation, the Schr\"{o}dinger equation can be written
\be 
\hat{\e} \psi(\rr) - \frac{Z e^2}{\kappa r} \psi(\rr) = \left(E + \frac{\Delta}{2} \right) \psi(\rr),
\label{eq:Schr}
\ee 
where $\psi(\rr)$ is the wavefunction in position space and $\hat{\e}$ is a $V$-dependent operator corresponding to the Mexican hat dispersion of the conduction band \cite{McCann2013epb} (illustrated in Fig.\ \ref{fig:results}a--c).  Eq.\ (\ref{eq:Schr}) can be written in momentum space
as
% by first expressing $\psi(\rr)$ in terms of its Fourier transform $\pt(\qq)$,
%\be 
%\psi(\rr) = \int \frac{d^2 q}{(2 \pi)^2} e^{i \qq \cdot \rr} \pt(\qq),
%\nonumber 
%\ee
%and then taking the integral $\int d^2 r e^{-i \kk \cdot \rr}$ of both sides.  This procedure gives
\be 
\e(k) \pt(k) - \int \frac{d^2 q}{(2 \pi)^2} \frac{2 \pi Z e^2}{\kappa |\kk - \qq|} \pt(q) = \left(E + \frac{\Delta}{2} \right)  \pt(k),
\label{eq:Schk}
\ee
where 
$\pt(\kk) = \int d^2r \exp[-i \kk \cdot \rr] \psi(\rr)$
is the Fourier transform of the electron wavefunction.
Here we have assumed that the (ground state) wavefunction is radially symmetric, so that $\pt(\kk) = \pt(k)$.

In the limit of asymptotically small ionization energy $|E|$, the momentum-space wavefunction $\pt(k)$ is strongly peaked around $k = k_0$.  In other words, only momentum states near the bottom of the band are used, and $\e(k)$ can be expanded around $k = k_0$ to give
\be 
\e(k) \simeq \frac{\Delta}{2} + \frac{\hbar^2}{2 m^*} (k - k_0)^2,
\label{eq:parabola}
\ee 
where
\be 
m^* = \frac{\tp (\tp^2 +V^2)^{3/2}}{2 v^2 (2 \tp^2 V + V^3)}.
\label{eq:m}
\ee 
Similarly, the integral in Eq.\ (\ref{eq:Schk}) can be simplified by noting that since $\psi(q)$ is appreciable only close to $q = k_0$, the integral is taken effectively over a thin ring in momentum space with radius $k_0$, and in that sense the integral in Eq.\ (\ref{eq:Schk}) reminds one of the expression for the electrostatic potential at a point $\kk$ produced by a coplanar charged ring with radius $k_0$.  For wave vectors $\kk$ with $|k - k_0| \ll k_0$, the ``potential" created by this ring is essentially the same as the potential created by a long, straight wire.  Thus, the integral in Eq.\ (\ref{eq:Schk}) can be written
\begin{eqnarray} 
\int \frac{d^2 q}{(2 \pi)^2} \frac{2 \pi Z e^2}{\kappa |\kk - \qq|} \pt(q) & \simeq & 
\frac{Z e^2}{2 \pi \kappa}\int dq \int dq_l \frac{\pt(q)}{\sqrt{q_l^2 + (q-k)^2} } \nonumber \\
& \simeq & \frac{Z e^2}{\pi\kappa } \int dq \pt(q) \ln \left( \frac{k_0}{|k - q|} \right),
\nonumber
\end{eqnarray}
where $q_l$ represents the distance along the length of the ``straight wire," and the integration over $q_l$ is truncated at $|q_l| = k_0$.

With these simplifications we can rewrite the Schr\"{o}dinger equation, Eq.\ (\ref{eq:Schk}), as 
\be
\begin{split}
\frac{\hbar^2 \delta_k^2}{2 m^*} \pt(\delta_k) - \frac{Z e^2}{\pi\kappa } \int d\delta_q  \ln \left( \frac{k_0}{|\delta_k - \delta_q|} \right) \pt(\delta_q)   \\ 
= E \pt(\delta_k),
\label{eq:Sch-radial}
\end{split}
\ee
where here we have introduced the notation $\delta_k = k - k_0$ and $\delta_q = q - k_0$.  
Written in the form of the Eq.\ (\ref{eq:Sch-radial}), the Schr\"{o}dinger equation is identical to that of the one-dimensional (1D) hydrogen atom \cite{Loudon1959odh}:
\be 
\frac{\hbar^2 k^2}{2 m} \pt(k) - \frac{Z e^2}{\pi \kappa} \int dq \ln \left(\frac{1/\lambda}{|k - q|} \right) \pt(q) = E \pt(k),
\label{eq:1DH}
\ee
where $m$ is the physical electron mass and $\lambda$ is some small-distance cutoff to the Coulomb potential.  (In the absence of such a cutoff, the ionization energy of the 1D hydrogen atom is logarithmically divergent \cite{Loudon1959odh}.)  

The possibility of mapping electron bound states with a 2D Mexican hat spectrum to equivalent 1D problems was previously pointed out by Chaplik and Magarill \cite{Chaplik2006bsi} in the context of Hamiltonians with spin-orbit coupling.  This equivalence can be seen as the consequence of the 1D ring of minima in the dispersion relation, which produces an effectively 1D-like density of states $dn/d\e \propto \e^{-1/2}$ near the band edge \cite{Chaplik2006bsi}.  In our case, the equivalence between Eqs.\ (\ref{eq:Sch-radial}) and Eq.\ (\ref{eq:1DH}) allows us to read off the answer for the energy directly\cite{Loudon1959odh}, which gives the result announced at the beginning, Eq.\ (\ref{eq:E}).  The energy is plotted in Fig.\ \ref{fig:results}d in the form $E + \Delta/2$ for one particular value of $Z \alpha$.

The corresponding wavefunction is given by $\pt(k) \propto [1 + b^2 \delta_k^2]^{-1}$, where $b = a/2 \ln(k_0 a)$, which corresponds to the 1D Fourier transform of the spatial wavefunction $\psi(x) \propto \exp[-|x|/b]$ for the 1D hydrogen atom\cite{Loudon1959odh}, with $k \rightarrow \delta_k$.  Taking the inverse Fourier transform of $\pt(k)$ (in two dimensions) gives the result announced in Eq.\ (\ref{eq:psi}).  Thus, the wavefunction describing the bound state in gapped BLG has an exponentially-decaying envelope, as in the normal hydrogen atom, but is modulated by a fast oscillation whose wavevector increases with the applied voltage.  This oscillation produces nodes of the electron density $|\psi(\rr)|^2$, as shown in Fig.\ \ref{fig:results}e.

Both the energy and the spatial structure of the wave function can in principle be measured experimentally by scanning tunneling microscopy \cite{Luican-Mayer2012stm, Li2013ell} on clean BLG samples with isolated impurities.  Further, the tunability of the impurity energy, and the correspondence between the energy and the spatial structure of the wave function, allows for these impurity levels to be used as benchmarks\cite{Martin2009nli} in studies of BLG using tunneling microscopy.  Our results also serve as an important ingredient for understanding situations with a relatively large impurity concentration, where long-ranged potential fluctuations can play an important role \cite{Deshpande2009mdp, Rutter2011mpi}.

Of course, the validity of our results, Eqs.\ (\ref{eq:E}) and (\ref{eq:psi}), relies on the assumption that the ionization energy $|E|$ is much smaller than the depth of the conduction band minimum, so that the approximation of Eq.\ (\ref{eq:parabola}) is valid.  This assumption breaks down at very small voltages $V/t \ll \sqrt{Z \alpha}$, where the conduction band maximum at $k = 0$ is sufficiently poorly developed that the conduction band can be described by a monotonically increasing dispersion relation, and the wavefunction loses its spatial oscillations.  In the limit of vanishingly small voltage the conduction band can be described by $\epsilon(k) \simeq \Delta/2 + \hbar^2k^2/2m^*_0$, where $m^*_0 = \tp/2v^2$, and the corresponding electron energy becomes $E \simeq - 2 m^*_0 Z^2 e^4/\hbar^2 = - Z^2 \alpha^2 t$.

More broadly, our description relies on the single-band approximation, which is accurate only when $|E| \ll \Delta$.  In the opposite limit, the ionization energy becomes larger than the band gap and the electron state is absorbed into the valence band.  By examining Eqs.\ (\ref{eq:Delta}) and (\ref{eq:E}) one can see that the condition $|E| \ll \Delta$ cannot be satisfied for any voltage unless $Z \alpha \ll 1$, a condition which brings to mind the problem of ``atomic collapse" in superheavy atoms \cite{Pomeranchuk1945els, Zeldovich1972ess} and, more recently, in graphene \cite{Fogler2007shc, Pereira2008sci, Shytov2007aca, Wang2012mdq, Wang2013oac, Gamayun2009scc}.  Unlike in those problems, however, in BLG atomic collapse can be induced even for small $Z \alpha$  by tuning the applied voltage.  In particular, at small $V/\tp$ the band gap becomes vanishingly small, so that the electron state crosses the gap when $V/\tp \lesssim (Z \alpha)^2$.  At large $V/\tp$, on the other hand, the band gap saturates, while the electron bound state grows deeper as $\ln^2(V/\tp)$.  Thus, the electron bound state again crosses the gap at some exponentially large voltage $V/\tp \gtrsim Z \alpha \exp[1/Z \alpha]$.  

These conditions on $V$ and $Z \alpha$ have important implications for experimental realizations of BLG-based transistors.  
Namely, they suggest that when Coulomb impurities are present in the substrate, the impurity levels they create can extend deep into the band gap except under the fairly strict condition $Z \alpha \ll 1$ and within a particular window of voltage.  Further, one can consider that while isolated coplanar impurities create bound states only at a particular energy, impurities that are displaced from the plane of the BLG create shallower bound states, so that in in principle Coulomb impurities distributed throughout a three-dimensional substrate can create impurity levels with any energy between zero and that of Eq.\ (\ref{eq:E}).  
This conclusion is in contrast with models of short-ranged disorder, which predict either a hard gap in the density of states for small impurity concentration \cite{Abergel2012cmm, Young2011cgb} or a mid-gap density of states that is exponentially small \cite{Mkhitaryan2008dit}.

In order to roughly estimate the conditions under which isolated Coulomb impurities do \emph{not} fill the band gap, one can take the result for the energy in the single-band approximation and equate it with $-\Delta/2$.  (When this equality is satisfied, positive Coulomb centers with a range of distances from the BLG plane can create levels which completely fill the upper half of the band gap, while negative Coulomb centers similarly distributed create levels that fill the lower half.)  As a comment on the expected accuracy of this approximation, we note that using the same approach for the massive Dirac spectrum gives $E + \Delta/2 = 0$ at $Z \alpha = 1/\sqrt{2}$, while the correct result for 2D is $Z \alpha = 1/2$ \cite{Ho1998dei}.  Our calculation can be expected to have a similar level of accuracy.

\begin{figure}[htb]
\centering
\includegraphics[width=0.48 \textwidth]{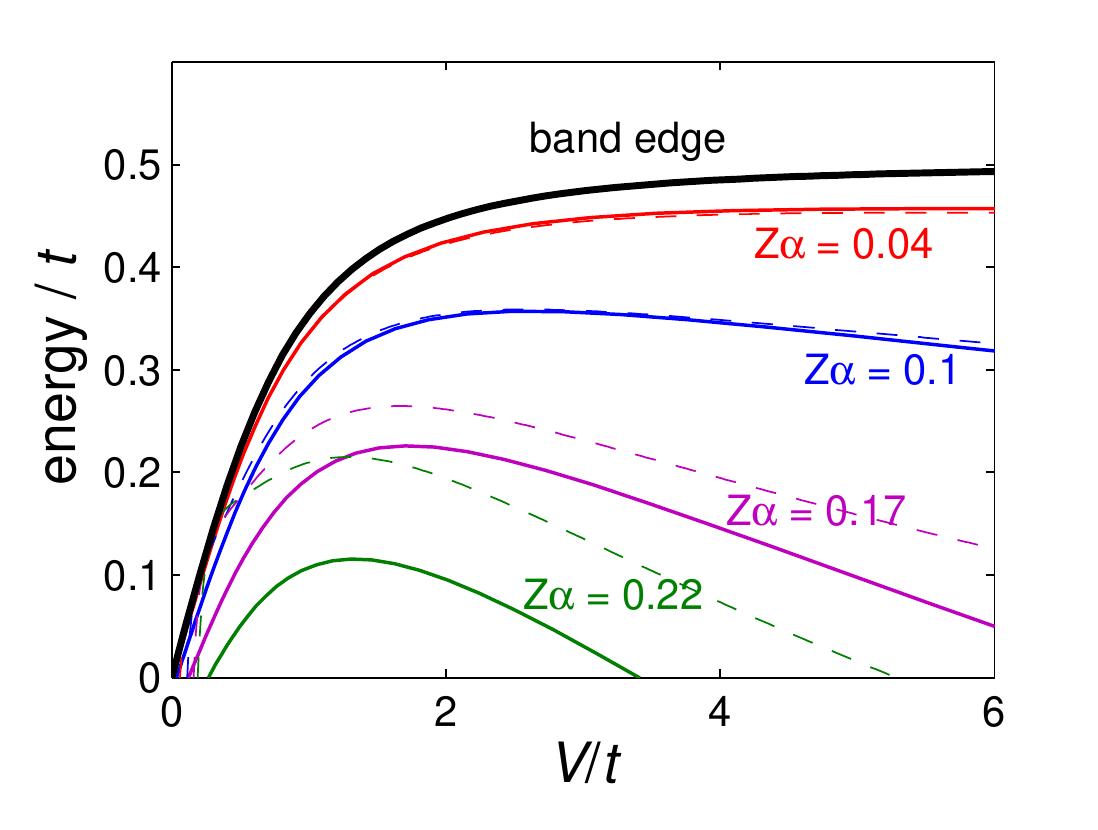}
\caption{(Color online) The energy of Coulomb impurity states, measured relative to mid-gap, as a function of $V$ for different values of $Z \alpha$.  Solid lines are calculated using a variational approach and the single-band approximation.  Dashed lines are the result of Eq.\ (\ref{eq:E}).  Different sets of curves are labeled by their corresponding value of $Z \alpha$.  At $Z \alpha \gtrsim 0.26$ energy levels sink below mid-gap for all values of $V$.}
\label{fig:different_alphas}
\end{figure}

Since the analytical result of Eq.\ (\ref{eq:E}) is accurate only at asymptotically small $Z \alpha$, a calculation at only moderately small $Z \alpha$, where $|E|$ becomes comparable to $\Delta/2$, must be done more carefully.  For this we use a variational approach, using as our variational wavefunction that of Eq.\ (\ref{eq:psi}) with the exponential decay length $b = a/2 \ln(k_0 a)$ kept as a variational parameter.  Inserting this wavefunction into Eq.\ (\ref{eq:Schk}) and optimizing $b$ numerically gives an improved single-band estimate for the energy
\footnote{A direct numerical solution of Eq.\ (\ref{eq:Schk}) gives an essentially identical result to that of the variational approach, to within our numerical accuracy.}.  
Our result is shown in Fig.\ \ref{fig:different_alphas} as a function of $V$ for a range of different values of $Z \alpha$.
For comparison is plotted the result of Eq.\ (\ref{eq:E}), which is shown to closely match the variational result when $Z \alpha$ is small and $V/\tp \gtrsim \sqrt{Z \alpha}$.

From our numerical calculation we can also estimate that the electron bound states cross below mid-gap for all $V$ when $Z \alpha \gtrsim 0.26$.  This result implies that observation of a hard gap that is not filled by Coulomb impurity levels requires a large dielectric constant $\kappa \gtrsim 8.5$ for monovalent impurities ($Z = 1$).  So far we are unaware of any experiments probing BLG on substrates with such large dielectric constant, although various high-$\kappa$ substrates have been explored for monolayer graphene\cite{Ponomarenko2009ehi}.  Another possibility is to place to the BLG in proximity to an additional, electrically isolated graphene layer with large electron density, which provides screening of the Coulomb potential\cite{Ponomarenko2011tmi}.  Again, our estimate for the required value of $Z \alpha$ comes within the single-band approximation, which is marginal when applied to the condition $E +\Delta/2 = 0$, and can be expected to be correct only to within a factor $\sqrt{2}$ or so.  An accurate description of energy levels close to the center of the band gap remains a challenge.

Finally, we note that in this paper we have ignored the role of dielectric polarization of the BLG.  Such polarization is notoriously strong in monolayer graphene on account of the gapless spectrum\cite{Ando2006sea, Goerbig2011epg, Skinner2013edr}, and leads to a renormalization of the fine structure constant toward lower values: $\alpha \rightarrow \alpha/(1 + \pi \alpha/2)$ \cite{Gorbar2002mfd}.  In BLG, however, the strong dielectric response\cite{Nandkishore2010dsa} can be gapped by the applied voltage, and is effectively eliminated for all wave vectors \cite{Kotov2008pcd} $q$ with $q \ll \Delta/\hbar v$.  The electron energy is therefore largely unaffected by dielectric response as long as $b \Delta/\hbar v \gg 1$.  This condition corresponds to $V \gg \sqrt{Z \alpha}$, which is already a condition for the validity of Eq.\ (\ref{eq:E}).  In the opposite regime, where the bound state energy closely approaches the mid-gap, excitonic effects are predicted to strongly modify the Coulomb interaction at short distances \cite{Cheianov2012gbg}, so that the impurity state can be qualitatively different from what we have described here.

We are grateful to V.\ Cheianov, A.\ V.\ Chaplik, V. Fal'ko, V. Gusynin, A.\ Kamenev, A.\ Luican-Mayer, K.\ A.\ Matveev, and A.\ F.\ Young for useful discussions.
This work was supported primarily by the MRSEC Program of the National Science Foundation under Award Number DMR-0819885. Work at Argonne National Laboratory supported by the U.S. Department of Energy, Office of Basic Energy Sciences under contract no. DE-AC02-06CH11357.
M.B.V. is supported, in part, by the DOE grant DE-FG02-94ER40823.

\bibliography{BLG_Coulomb_boundstate}

\end{document}